\documentclass[a4paper,11pt]{article}
\pdfoutput=1 

\usepackage{jheppub} 

\usepackage[T1]{fontenc} 

\title{\boldmath The electromagnetic form factors of heavy-light pseudo-scalar and vector mesons}

\author[a,b,1]{Yin-Zhen Xu}

\affiliation{Departmento de Ciencias Integradas, Universidad de Huelva, E-21071 Huelva, Spain.}
\affiliation{Departamento de Sistemas F\'isicos, Qu\'imicos y Naturales, Universidad Pablo de Olavide, E-41013 Sevilla, Spain}
\emailAdd{yinzhen.xu@dci.uhu.es}

\abstract{We systematically investigate the electromagnetic form factors of heavy-light pseudo-scalar and vector mesons within the Dyson-Schwinger/Bethe-Salpeter equations framework for the first time. It is found that the charge radius of vector meson is larger than that of its pseudo-scalar counterpart. In heavy-light systems, the flavor symmetry breaking will lead to a splitting of the form factor of different quark, and the distribution range of lighter and heavier quark gradually expands and contracts, respectively. The competition between them together generates the electromagnetic form factors of meson. Our results can be compared with other theoretical calculations and future experimental data.
}

\begin{document} 
\maketitle
\flushbottom

\section{Introduction}
The study of electromagnetic properties of mesons is a fundamental topic in hadron physics. The electromagnetic form factors (EFFs), which describe the response of composite particles to electromagnetic probes, provide an important tool for understanding the structure of bound states in QCD. Therefore, 
there have been numerous studies of the electromagnetic form factors of the mesons in experimental and theoretical \cite{Belle-II:2018jsg,Aguilar:2019teb,Horn:2017csb,Bijnens:2002hp,Boyle:2008yd,Kwee:2007dd,Gao:2017mmp,Miramontes:2022uyi,Sauli:2022ild}. \par
 Compared with $u\bar{d}$, $u\bar{s}$, ..., heavy-light systems, such as $u\bar{c}$, $u\bar{b}$, $c\bar{s}$, ..., exhibit higher flavor asymmetry, thereby offering more information for the internal structure and dynamics of QCD's bound states. However, due to the lack of experimental data and theoretical challenges, there are few studies about them. Recently, the electromagnetic form factors of heavy-light meson have attracted growing attention and various methods have been applied, for example, light-front framework (LFF) \cite{Hwang:2001th}, constituent quark model (CQM) \cite{Moita:2021xcd}, contact interaction model (CI) \cite{Hernandez-Pinto:2023yin}, Algebraic model (AM) \cite{Almeida-Zamora:2023bqb}, Extended Nambu–Jona-Lasinio model (ENJL) \cite{Luan:2015goa}, Lattice QCD (lQCD) \cite{Li:2017eic} and others \cite{Das:2016rio,Aliev:2019lsd,Simonis:2016pnh,Bose:1980vy,Lahde:2002wj,Wang:2019mhm}. Different or similar results have been reported. \par 
On the other hand, the Dyson-Schwinger/Bethe-Salpeter equations (DSEs/BSEs) formalism  provides a non-perturbative and Poincaré-covariant framework capable of simultaneously describing confinement and dynamical chiral symmetry breaking (DCSB). It has been successfully used to study hadron properties for thirty years \cite{Roberts:1994dr,Maris:1999nt,Maris:2003vk,Maris:1997tm,Qin:2011dd,Rodriguez-Quintero:2010qad,Qin:2011xq,Xu:2020loz,Qin:2019oar,Xu:2021lxa,Li:2023zag,daSilveira:2022pte}. Therefore, investigating the electromagnetic form factors of heavy-light systems within the framework of DSEs/BSEs, including the extraction of physical information like charge radius and magnetic moment, is necessary and valuable for comparative analysis with results obtained from alternative approaches and future experiments. However, the current predictions of DSEs/BSEs for the meson's EFFs have focused on flavor-symmetric or slightly asymmetric systems such as $\pi$, $\rho$, and $K$ mesons \cite{Maris:2000sk,Bhagwat:2006pu,Chen:2018rwz,Xu:2019ilh,Xu:2023vlt,Xu:2023izo}. \par
In this work, we extend our previous work \cite{Xu:2019ilh} to the heavy-light mesons, and an effective flavor-dependent BSE interaction kernel is applied \cite{Qin:2019oar}. Based on it, we systematically investigate the electromagnetic form factors of heavy-light quark-antiquark system within DSEs/BSEs framework for the first time. As a comparison, pseudo-scalar (PS) and vector (VC) channels are calculated uniformly.\par  
This paper is organized as follows: In section. \ref{sec:2}, we introduce the DSEs/BSEs framework and the electromagnetic form factors of meson. In section. \ref{sec:3}, the numerical results of heavy-light pseudo-scalar/vector mesons' EFFs are presented. Then we discuss the effect of flavor symmetry breaking, and the results are compared with those obtained by other approaches. Section. \ref{sec:4} provides a brief summary.\par 
\section{The electromagnetic form factors within DSEs/BSEs framework}
\label{sec:2}
\subsection{Quark propagators, quark-photon vertex and heavy-light meson's BSAs}
We work within DSEs/BSEs framework in Euclidean space. Under the rainbow-ladder (RL) approximation, the dressed-quark propagator can be obtained from the following gap equation,
\begin{equation}
  S^{-1}(p)=Z_2i \gamma \cdot p + Z_4 m + Z_{1} \int_q^{\Lambda} g^2 D_{\alpha \beta}(p-q) \frac{\lambda^a}{2} \gamma_\alpha S(q) \frac{\lambda^a}{2} \gamma_\beta,
  \label{eq.DSE}
\end{equation} 
and the general form of $S^{-1}(p)$ can be written as
\begin{equation}
  S^{-1}(p)= i\gamma \cdot pA(p^2) + B(p^2).
\end{equation}
Where $A(p^2)$ and $B(p^2)$ are scalar functions, $m$ is current quark mass, $Z_{1,2,4}$ are the renormalization constants, in this work we employ a mass-independent momentum-subtraction renormalisation scheme \cite{Chang:2008ec} and choose renormalization scale $\xi = 19$ GeV \cite{Xu:2021mju}. $\int_q^{\Lambda}$ represents a translationally-invariant regularization of the four-dimensional integral with the regularization scale $\Lambda$. For the dressed-gluon propagator $D_{\mu\nu}(k)$, we employ the Qin-Chang model \cite{Qin:2011dd,Qin:2011xq}
\begin{equation}
  Z_1 g^2 D_{\mu \nu}(k)=Z_2^2 \mathcal{G}\left(k^2\right) \mathcal{P}_{\mu \nu}^T(k)=\mathcal{D}_{\text{eff}},
  \label{eq.gluon}
\end{equation}
where $\mathcal{P}_{\mu \nu}^T(k)=\delta_{\mu \nu}-{k_\mu k_\nu}/k^2$ is transverse projection operator and the effective interaction is
\begin{equation}
\frac{\mathcal{G}(k^2)}{k^2}=\mathcal{G}^{\text{IR}}(k^2)+\frac{8\pi^2\gamma_{m}\mathcal{F}(k^2)}{\ln[\tau+(1+k^2/\Lambda^2_{\text{QCD}})^2]},\ \mathcal{G}^{\text{IR}}(k^2)=D\frac{8\pi^2}{\omega^4}e^{-k^2/\omega^2},
\label{eq.gluon}
\end{equation}
with $\mathcal{F}(k^2)=\{1-\exp[(-k^2/(4m_t^2)]\}/k^2$, $m_t=0.5$\,GeV, $\tau=e^2-1$, $\Lambda_{\text{QCD}}=0.234$\,GeV, $\gamma_{m}=12/25$ \cite{Xu:2021mju}. For model parameters $D$ and $\omega$, a typical choice is $\omega=0.5$ GeV with $D\omega=(0.82\ \text{GeV})^3$ for $u/d$, $s$ quark, $\omega=0.8$ GeV with $D\omega=(0.6\ \text{GeV})^3$ for $c$, $b$ quark \cite{Xu:2021mju,Xu:2019ilh}. In this work we follow these values except $(D\omega)_s=(0.68\ \text{GeV})^3$, $(D\omega)_c=(0.66\ \text{GeV})^3$ ,$(D\omega)_b=(0.48\ \text{GeV})^3$, which are tweaked slightly to consider flavor dependence of the interaction \cite{Serna:2017nlr,daSilveira:2022pte,Chen:2019otg,Serna:2020txe}, thus producing results closer to the experimental value (see Table. \ref{tab:1}). More details of Eq. (\ref{eq.DSE}-\ref{eq.gluon}) are presented in Refs. \cite{Roberts:1994dr,Maris:2003vk,Maris:1997tm,Qin:2011dd,Maris:1999nt}.\par 
\begin{table}
\centering
\scalebox{0.8}{
\begin{tabular}{|c|c|c|lll|lll|}
\hline
 Quark & $m^{\xi=19\ \text{GeV}}$& Meson &\multicolumn{3}{|c|}{Mass} &\multicolumn{3}{c|}{Decay constant}\\
  &  &  &  Expt. & lQCD & This work  &Expt. & lQCD & This work \\ \hline
$u/d$ & 0.0033 & $\pi$ & 0.138(1) & - & 0.135  & 0.092(1) & 0.093(1) & 0.095\\
         & /  & $\rho$ & 0.775(1) & 0.780(16) & 0.755 & 0.153(1)& - & 0.150\\
$s$  & 0.097 & $\eta_s$ & - & - & 0.734 & - & - & 0.123 \\
       & /  & $\phi$ & 1.019(1) & 1.032(16) & 1.019  & 0.168(1) & 0.170(13) & 0.168\\
$c$   & 0.854 & $\eta_c$ & 2.984(1) & - & 2.984 & 0.237(52) &0.278(2) &  0.270\\
    & / & $J/\psi$ & 3.097(1) & 3.098(3) & 3.114 & 0.294(5) & 0.286(4) & 0.290\\
$b$  & 3.682 & $\eta_b$ & 9.399(1) & - & 9.399 & - & 0.472(5) & 0.464\\
     & / & $\Upsilon$ & 9.460(1) & - & 9.453 & 0.505(4) & 0.459(22) & 0.441\\
\hline
\end{tabular}
}
\caption{\label{tab:1} The masses and decay constants of $q\bar{q}$ systems. Where $m$ is current quark mass and the units are GeV. For comparison, we collect both experimental values \cite{ParticleDataGroup:2018ovx} and lQCD's results \cite{Fu:2016itp,Donald:2013pea,Donald:2012ga,Follana:2007uv,McNeile:2012qf,Colquhoun:2014ica}. }
\end{table}

Correspondingly, the meson's Bethe-Salpeter amplitudes (BSAs), $\Gamma^{f\bar{g}}_H\left(k_{+}, k_{-}\right)$, and the dressed quark-photon vertex, $ \Gamma^{\gamma,f\bar{f}}_\mu\left(k_{+}, k_{-}\right) $, can be obtained from (in)homogeneous BSEs, respectively,
\begin{equation}
  \Gamma^{f\bar{g}}_H\left(k_{+}, k_{-}\right) = \int_{q}^{\Lambda} K^{f\bar{g}}(q,k;P)  S^{f}\left(q_{+}\right) \Gamma^{f\bar{g}}_H\left(q_{+}, q_{-}\right) S^{g}\left(q_{-}\right),
    \label{eq:hBSE}
\end{equation}
and
\begin{equation}
 \Gamma^{\gamma,f\bar{f}}_\mu\left(k_{+}, k_{-}\right) = Z_2 \gamma_\mu - \int_{q}^{\Lambda}  K^{f\bar{f}}(q,k;P)  S^{f}\left(q_{+}\right) \Gamma^{\gamma,f\bar{f}}_\mu\left(q_{+}, q_{-}\right) S^{\bar{f}}\left(q_{-}\right),
  \label{eq:iBSE}
\end{equation}
where $f$ and $g$ denote the flavor of (anti-)quark, $k_+= k + \alpha P$; $k_-= k - (1-\alpha) P$ with the momentum partitioning parameter $\alpha \in [0,1]$. Although the physical observables do not depend on $\alpha$, in the actual calculation\footnote{If we define the vertex of parabola contour in $k^2_\pm$ complex plane as $(- \Delta_{f} ^2, 0)$ and $(-\Delta_{\bar{g}}^2, 0)$ , we will have $1-\Delta_{\bar{g}}/M<\alpha<\Delta_f/M$, and the best $\alpha = \Delta_f / (\Delta_f + \Delta_{\bar{g}})$, where $M$ is meson's mass.}, the selection of $\alpha$ should be careful to avoid parabola include the pole for the accuracy of the contour integral \cite{Fischer:2005en,Blank:2011qk}. The general form of $\Gamma(k_+,k_-)$ can be written as 
\begin{equation}
  \Gamma (k_+, k_-)=\sum_{i=1}^N \tau^i(k, P) \mathcal{F}_i (k, P),
\end{equation}
where $\tau^i(k,P)$ is basis and $\mathcal{F}_i(k,P)$ is scalar function. For the pseudo-scalar/vector meson, we choose \cite{Qin:2011xq}
\begin{subequations}
\label{eq:basis}
\begin{align}
\tau_{0^{-}}^1&=i \gamma_5,& &\tau_{0^{-}}^3=\gamma_5 \gamma \cdot k k \cdot P,\nonumber \\
\tau_{0^{-}}^2&=\gamma_5 \gamma \cdot P,& &\tau_{0^{-}}^4=\gamma_5 \sigma_{\mu \nu} k_\mu P_\nu,
\end{align}
and
\begin{align}
& \tau_{1^{-}}^1=i \gamma_\mu^T, & &\tau_{1^{-}}^5=k_\mu^T,\nonumber  \\
& \tau_{1^{-}}^2=i\left[3 k_\mu^T \gamma \cdot k^T-\gamma_\mu^T k^T \cdot k^T\right], & &\tau_{1^{-}}^6=k \cdot P\left[\gamma_\mu^T \gamma^T \cdot k-\gamma \cdot k^T \gamma_\mu^T\right],\nonumber  \\
& \tau_{1^{-}}^3=i k_\mu^T k \cdot P \gamma \cdot P,& & \tau_{1^{-}}^7=\left(k^T\right)^2\left(\gamma_\mu^T \gamma \cdot P-\gamma \cdot P \gamma_\mu^T\right)-2 k_\mu^T \gamma \cdot k^T \gamma \cdot P, \nonumber \\
& \tau_{1^{-}}^4=i\left[\gamma_\mu^T \gamma \cdot P \gamma \cdot k^T+k_\mu^T \gamma \cdot P\right], & &\tau_{1^{-}}^8=k_\mu^T \gamma \cdot k^T \gamma \cdot P,
\end{align}
\end{subequations}
with $V_\mu^T=V_\mu-P_\mu(V \cdot P) / P^2$. Then the decay constant can be obtained easily after normalization of the meson' BSAs \cite{Qin:2011xq}.\par 
As for the quark-photon vertex $\Gamma^\gamma_\mu (k_+,k_-)$, the general basis is
\begin{equation}
\tau_\gamma = \{\gamma_\mu,k_\mu, P_\mu \}\otimes \{ 1,\ \gamma\cdot P,\ \gamma \cdot k,\ \sigma_{\alpha\beta}P^{\alpha}k^{\beta} \},
\end{equation}
and $\Gamma^\gamma_\mu (k_+,k_-)$ should satisfy the vector Ward-Green-Takahashi identity (WGTI) \cite{Maris:2000sk,Xu:2023izo}
\begin{equation}
\label{eq:WGTI}
i P_\mu \Gamma^\gamma_\mu \left(k_+, k_-\right)=S^{-1}\left(k_+\right)-S^{-1}\left(k_-\right).
\end{equation}\par 
The only thing left is the BSE interaction kernel $K^{f\bar{g}}(q,k;P)$. In the case of flavor symmetry, the standard RL approximation kernel can be written as \cite{Xu:2021mju}
\begin{equation}
\label{eq:kernel.RL}
  K^{f\bar{f}}_{\text{RL}}(q,k;P) = \mathcal{D}^f_{\text{eff}}\frac{\lambda^a}{2} \gamma_\alpha \otimes \frac{\lambda^a}{2} \gamma_\beta,
\end{equation}
and generally, it works well for the flavor symmetric/slightly asymmetric ground state pseudo-scalar/vector mesons. However, for the highly flavor asymmetric system, such as $u\bar{c}$, $u\bar{b}$, $\dots$, Eq. (\ref{eq:kernel.RL}) is difficult to be applied because of the lack of flavor asymmetry \cite{Chen:2019otg}. To consider this effect, we effectively average the kernel as \cite{Qin:2019oar}
\begin{equation}
\label{eq:kernel.wRL}
  K^{f\bar{g}}(q,k;P) = \eta K_{\text{RL}}^{f\bar{f}}(q,k;P) + (1-\eta) K_{\text{RL}}^{g\bar{g}}(q,k;P). 
  \end{equation}
Here a weight factor $\eta$ is introduced \footnote{In this work, "$\eta$" is just used to denote the weight factor, and it has nothing to do with the momentum partition parameter $\alpha$ defined below Eq. (\ref{eq:iBSE}).}, and Eq. (\ref{eq:kernel.wRL}) can be regarded as an extension of RL approximation, that is, weight-RL. For the flavor symmetric case, it will degenerate to the RL kernel, therefore the solution of Eq. (\ref{eq:iBSE}) still satisfy the vector WGTI, which will be used in the calculation of electromagnetic form factor \cite{Maris:2000sk}. \par 
\begin{figure}
\centering 
\includegraphics[width=0.48\textwidth]{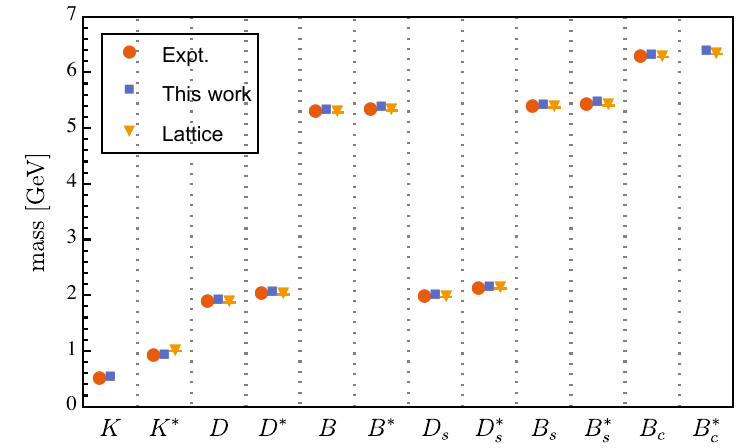}
\includegraphics[width=0.48\textwidth]{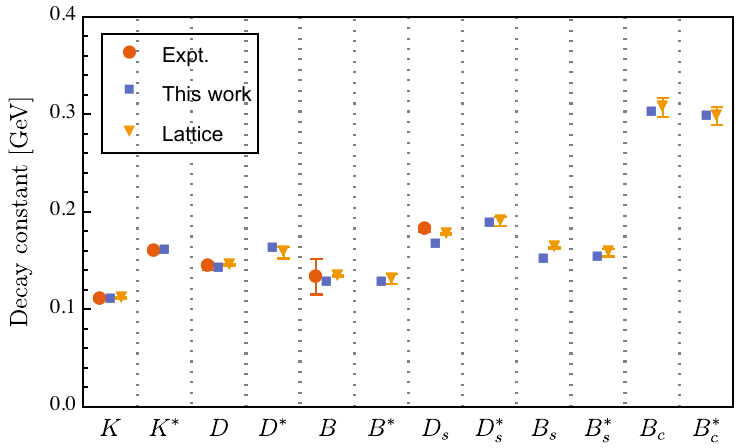}
\caption{\label{fig:mf} The masses and decay constants of flavor asymmetric pseudo-scalar/vector mesons. See Table. \ref{tab:2} for more details.}
\end{figure}
For flavor-asymmetric mesons, the effect of weight factor has been discussed in Ref. \cite{Qin:2019oar}, with the presentation of an automatic average. In this work, we directly determine the $\eta$ by the pseudo-scalar meson's mass to obtain a relatively realistic interaction. Once the weight factor is fixed, the decay constant of the pseudo-scalar meson, the mass and decay constant of the vector meson can all be well predicted (see Table. \ref{tab:2} and Figure. \ref{fig:mf}). It is worth noting that Eq. (\ref{eq:kernel.wRL}) should be considered as an effective kernel and other heavy-light kernel have been proposed. For example, the modification of the combined effect of vertex and gluon dressing, suppressed for heavier quarks, was first introduced by Serna et al in contact-interaction model to study heavy-light meson \cite{Serna:2017nlr}, and subsequently applied in the DSEs/BSEs framework with a kernel ansatz \cite{Chen:2019otg,Serna:2020txe}. The theoretical explore for flavor dependence and strict beyond-RL kernel is still ongoing \cite{Qin:2019oar,Ivanov:1998ms,Rojas:2014aka,Gomez-Rocha:2016cji,daSilveira:2022pte,Serna:2020txe,Serna:2017nlr,Chen:2019otg,Qin:2020jig}.
\begin{table}
\centering
\scalebox{0.8}{
\begin{tabular}{|l|llll|llll|}
\hline
 Meson&\multicolumn{4}{|c|}{Mass} &\multicolumn{4}{c|}{Decay constant}\\
  & Expt. & lQCD & This work & RL  &Expt. & lQCD & This work & RL\\ \hline
 $K$ & 0.495(1) &- & 0.495$^\dagger$&0.495 &0.110(1)&-& 0.108& 0.112\\
 $K^*$ & 0.896(1)& 0.993(1) & 0.880 & 0.955 &0.159(1)&-&0.158&0.179 \\
 $D$&1.868(1)&1.868(3)& 1.868$^\dagger$ & - & 0.144(4)&0.150(4)& 0.140& -  \\
 $D^*$&2.009(1)&2.013(14)& 2.017 & - &-& 0.158(6)&0.160& -  \\
 $B$&5.279(1)&5.283(8)&5.279$^\dagger$ &-&0.133(18)&0.134(1)&0.123& - \\
 $B^*$&5.325(1)&5.321(8)&5.334&-&-&0.131(5)&0.126& - \\
 $D_s$&1.968(1)&1.968(4)&1.968$^\dagger$&-&0.182(3)&0.177(1)&0.164& - \\
 $D^*_s$&2.112(1)&2.116(11)&2.111&-&-&0.190(5)&0.186& - \\
 $B_s$&5.367(1)&5.366(8)&5.367$^\dagger$&-&-&0.163(1)&0.149& - \\
 $B^*_s$&5.415(1)&5.412(6)&5.422&-&-&0.158(4)&0.151& - \\
 $B_c$&6.275(1)&6.276(7)&6.275$^\dagger$&6.388&- &0.307(10)&0.300& 0.429 \\
 $B^*_c$&-&6.331(7)&6.340&6.542&-&0.298(9)&0.296& 0.483 \\
  \hline
\end{tabular}
}
\caption{\label{tab:2} The masses and decay constants of heavy-light mesons, where $^\dagger$ means fitting value. In this work (weight-RL), we fixed the weight factor $\eta$ by the mass of the pseudo-scalar meson (see Eq. (\ref{eq:kernel.wRL})): $\eta_{u\bar{s}}=0.467$,\ $\eta_{u\bar{c}}=0.311$,\ $\eta_{u\bar{b}}=0.249$,\ $\eta_{s\bar{c}}=0.410$,\ $\eta_{s\bar{b}}=0.325$,\ $\eta_{c\bar{b}}=0.415$. For comparison, we collect both experimental values \cite{ParticleDataGroup:2018ovx,CLEO:2000moj} and lQCD's results \cite{Mathur:2018epb,Cichy:2016bci,Dowdall:2012ab,Fu:2016itp,Dudek:2014qha,Donald:2013pea,Lubicz:2017asp,Donald:2012ga,Follana:2007uv,McNeile:2012qf,Bazavov:2017lyh,Colquhoun:2014ica,Colquhoun:2015oha}, the RL results are picked from Refs. \cite{Xu:2019ilh,Qin:2019oar}. The units in this table are GeV.}
\end{table}

\subsection{The electromagnetic form factors of pseudo-scalar and vector mesons}

The generalized impulse approximation allows electromagnetic processes to be described in terms of dressed quark propagators, bound state BSAs, and the dressed quark-photon vertex. These couplings are given by \cite{Maris:2000sk,Bhagwat:2006pu,Xu:2019ilh}
\begin{equation} 
\Lambda^{\mu,\bar{f}gg}_{H}(P, Q) = iN_{c} \int_{k} \operatorname{Tr} \Big[ \Gamma^{\gamma}_{\mu}(k_+;k_-) S^g(k_-) \Gamma^{(\text{in})}_H(k_-;k_p) S^{\bar{f}}(k_p)\Gamma^{(\text{out})}_H(k_-;k_p) S^g(k_+) \Big].
\label{eq.QPV}
\end{equation}
Where $P-Q/2$, $P+Q/2$ and $Q$ are incoming meson, outgoing meson and incoming photon momenta, which are constrained by on-shell condition
\begin{equation}
(P-Q/2)^2=(P+Q/2)^2=-M^2,
\end{equation}
with $M$ is meson's mass. The other elements in Eq. (\ref{eq.QPV}) are the dressed-quark propagators $S(q)$, the meson's BSAs $\Gamma_H(k_+,k_-)$ and dressed quark-photon vertex $\Gamma^\gamma_\mu(k_+,k_-)$. In this work we determine them from homogeneous/inhomogeneous BSEs in the moving frame, correspondingly, $k_+=k+(1-\alpha) P+Q/2$, $k_-=k+(1-\alpha)P-Q/2$, $k_p=k-\alpha P$, with $\alpha$ is the momentum partitioning parameter of the outgoing meson and physical observables are independent of it. Then the need for interpolation or extrapolation of the meson's BSAs/quark-photon vertex can be avoided \cite{Bhagwat:2006pu}.  \par
\begin{figure}
\centering 
\includegraphics[width=1\textwidth]{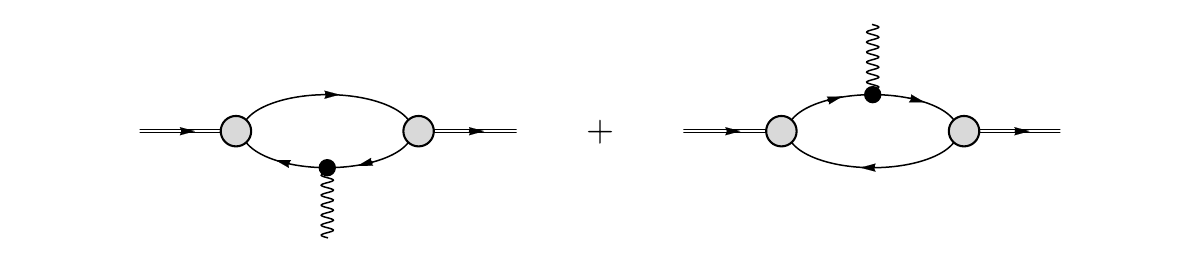}
\caption{\label{fig:fey2}Feynman diagrams of meson's electromagnetic form factor under impulse approximation. Solid lines: dressed quarks, $S$; gray shaded circles: the Bethe-Salpeter amplitude of meson, $\Gamma_H$; and black shaded circles: dressed quark-photon vertex, $\Gamma^\gamma_\mu$.}
\end{figure}

 
Consider the coupling of a photon to the quark and antiquark, this interaction should be written as the sum of two terms (see Figure. \ref{fig:fey2})

\begin{equation}
\label{eq:eff}
  \Lambda^{\mu}_{H}(P, Q) = \hat{Q}^g \Lambda^{\mu,\bar{f}gg}_{H}(P, Q) + \hat{Q}^{\bar{f}} \Lambda^{\mu,g\bar{f}\bar{f}}_{H}(P, Q),
\end{equation}
where $\hat{Q}$ is the quark or antiquark electric charge. Therefore, the meson's form factor can be decomposed into two terms 
\begin{align}
  F^{g\bar{f}}(Q^2) = \hat{Q}^g  F^{g}_{g\bar{f}}(Q^2) + \hat{Q}^{\bar{f}}  F^{\bar{f}}_{g\bar{f}}(Q^2),
  \label{eq.eff2}
\end{align}
where $F^{g}_{g\bar{f}}(Q^2)$ and $F^{\bar{f}}_{g\bar{f}}(Q^2)$ describe the contribution of different (anti-)quark in the system, respectively.\par 
For pseudo-scalar meson, the only form factor is defined by \cite{Maris:2000sk}
\begin{equation}
F(Q^2) = \frac{P^\mu}{2P^2}\Lambda^{\mu}(P, Q),
\end{equation}
and for vector meson, the three form factors are defined by \cite{Bhagwat:2006pu}
\begin{subequations}
\begin{align}
G_E\left(Q^2\right)&=\left(1+\frac{2}{3}\frac{Q^2}{4 M^2}\right)  F_1\left(Q^2\right)+ \frac{2}{3} \frac{Q^2}{4 M^2}  F_2\left(Q^2\right) +\frac{2}{3} \frac{Q^2}{4 M^2} \left(1+\frac{Q^2}{4 M^2}\right) F_3\left(Q^2\right), \\
G_M\left(Q^2\right)&=-F_2\left(Q^2\right), \\
G_\mathcal{Q}\left(Q^2\right) &= F_1\left(Q^2\right) + F_2\left(Q^2\right) + \left(1+\frac{Q^2}{4 M^2}\right) F_3\left(Q^2\right),
\end{align}
\end{subequations}
with
\begin{subequations}
\begin{align}
\Lambda^{\mu}_{\rho \sigma}(P, Q)&=-\sum_{j=1}^3 T_{\mu \rho \sigma}^{j}(P, Q) F_j\left(Q^2\right),\\
T_{\rho \sigma}^{\mu,1}(P, Q)&=  2 P_\mu \mathcal{P}_{\rho \gamma}^T\left(P^{-}\right) \mathcal{P}_{\gamma \sigma}^T\left(P^{+}\right), \\
T_{\rho \sigma}^{\mu,2}(P, Q)&=\left(Q_\rho-P_\rho^{-} \frac{Q^2}{2 M^2}\right) \mathcal{P}_{\mu \sigma}^T\left(P^{+}\right) -\left(Q_\sigma+P_\sigma^{+} \frac{Q^2}{2 M^2}\right) \mathcal{P}_{\mu \rho}^T\left(P^{-}\right), \\
T_{\rho \sigma}^{\mu,3}(P, Q)&=\frac{P_\mu}{M^2}\left(Q_\rho-P_\rho^{-} \frac{Q^2}{2 M^2}\right) \left(Q_\sigma+P_\sigma^{+} \frac{Q^2}{2 M^2}\right),
\end{align}
\end{subequations}
and  $\mathcal{P}_{\mu \nu}^T(k)=\delta_{\mu \nu}-{k_\mu k_\nu}/k^2$. Where $G_E(0)=F(0)=e$, $e=1,0$ defines the meson's charge. In the impulse approximation, as long as the relation between the dressed quark propagator and the quark-photon vertex satisfies vector WGTI, and the BSE kernel is independent of the meson momentum, the conservation of electromagnetic current will be preserved after the meson's BSAs are canonical normalized \cite{Maris:2000sk}. Besides, $G_{M}$(0) and $G_{\mathcal{Q}}(0)$ can be identified with the magnetic moment, $\mu$, and the quadrupole moment, $\mathcal{Q}$, of a vector meson \cite{Bhagwat:2006pu,Xu:2019ilh}.

\section{Numerical results and discussion}
\label{sec:3}
\subsection{Pseudo-scalar mesons}
\begin{figure}
\centering 
\includegraphics[width=.49\textwidth]{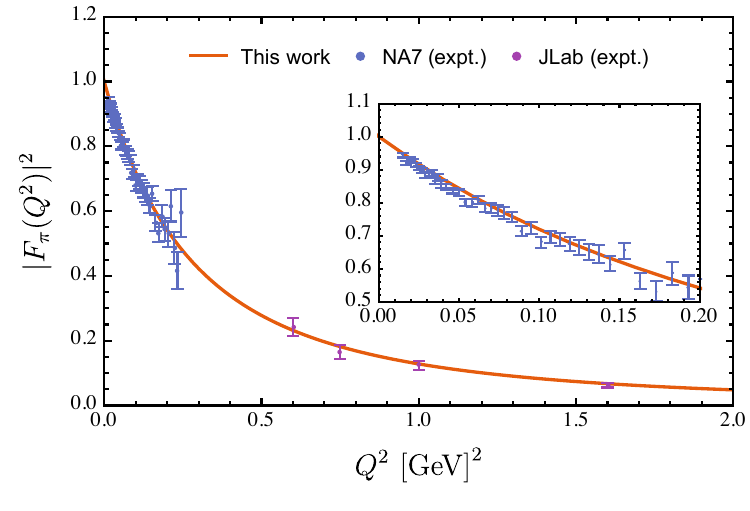}
\includegraphics[width=.49\textwidth]{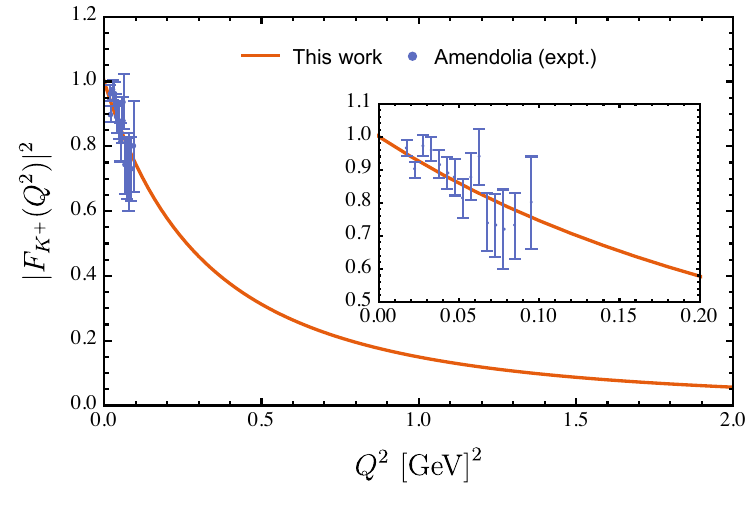}
\caption{\label{fig:pk}Compare the electromagnetic form factors of pion and kaon with experimental data \cite{NA7:1986vav,JeffersonLabFpi:2000nlc,Amendolia:1986ui}.}
\end{figure}

With all the above at hand, we first consider the electromagnetic form factor for ground state heavy-light pseudo-scalar mesons. These are the simplest quark-antiquark bound states embodying confinement, and the lightest pseudo-scalar meson, $\pi$, is also the Goldstone mode of DCSB. In Figure \ref{fig:pk}, we present a comparison of the electromagnetic form factors of the pion and kaon with current experimental results. Our predictions demonstrate a good agreement with the experimental data. \par 
In the case of pseudo-scalar channel, the only electromagnetic form factor, $F(Q^2)$, corresponds to the charge distribution of the system\footnote{Although the exact form of this relation is still up for debate \cite{Epelbaum:2022fjc,Miller:2018ybm}.}. It is well known that the charge radius can be defined as
\begin{equation}
\label{eq.r}
  \left\langle r^2\right\rangle=-\left. 6\frac{d F\left(Q^2\right)}{d Q^2}\right|_{Q^2=0},
\end{equation}
which denotes the distribution range of charge. The numerical results of Eq. (\ref{eq.r}) can be found in Table. \ref{tab:PS}. As a comparison, we also collected some predictions from other approaches.
\begin{table}
\centering
\scalebox{0.8}{
\begin{tabular}{|l|l|lll|l|l|l|l|l|l|l|}
\hline
 \multicolumn{2}{|c}{Meson}&\multicolumn{3}{|c|}{This work} & LFF & PM & CI & AM &CQM&ENJL&IQCD \\
  \multicolumn{2}{|c|}{}  & \multicolumn{1}{c}{L} & \multicolumn{1}{c}{H} & \multicolumn{1}{c|}{Full} & &  & & & & & \\ \hline
 $\pi $&$u\bar{d}$ & 0.646 & 0.646 & 0.646 & 0.666 & - & 0.45 & - & 0.665& 0.57& 0.648(15) \\ 
 $K^+$&$u\bar{s}$ & 0.659 & 0.491 & 0.608 & 0.591 & - & 0.42 & -&0.551&0.54 & - \\
 $K^0$&$d\bar{s}$ & / & / & 0.253$i$ & 0.260$i$ & -&- & -& - &- &- \\
 $D^+$&$c\bar{d}$ & 0.706 & 0.187 & 0.435 & 0.429 & 0.510 & - & 0.680& 0.505 &0.46 &0.450(24) \\
 $D^0$&$c\bar{u}$ & / & /& 0.556$i$ & 0.551$i$ & 0.673$i$ & 0.36$i$ & -& - &- &- \\
 $B^+$&$u\bar{b}$ & 0.757& 0.071 & 0.619 & 0.615 & 0.732 & 0.34 &0.926& - &0.74 &- \\
 $B^0$&$d\bar{b}$ & / & / & 0.435$i$ & 0.432$i$ & 0.516$i$ & - & -& -&- &- \\
 $D_s$&$c\bar{s}$ & 0.547 & 0.192 & 0.352 & 0.352 & 0.465 & 0.26 & 0.372&0.377 &0.39 &0.465(57) \\
 $B_s$&$s\bar{b}$ & 0.588 & 0.072 & 0.337$i$ & 0.345$i$ & 0.463$i$& 0.24$i$ & 0.345$i$& -&- &-\\
 $B_c$&$c\bar{b}$ & 0.260 & 0.089 & 0.219 & 0.208 & - & 0.17 & 0.217 &- &- &- \\
  \hline
\end{tabular}
}
\caption{\label{tab:PS} The charge radius $\sqrt{\langle r^2 \rangle}$ of heavy-light pseudo-scalar mesons, including the contributions of lighter quark (L) and heavier quark (H). The results of other approaches come from: light-front framework (LFF) \cite{Hwang:2001th}, QCD potential model (PM) \cite{Das:2016rio}, constituent quark model (CQM) \cite{Moita:2021xcd}, contact interaction model (CI) \cite{Hernandez-Pinto:2023yin}, Algebraic model (AM) \cite{Almeida-Zamora:2023bqb}, Extended Nambu–Jona-Lasinio model (ENJL) \cite{Luan:2015goa} and Lattice QCD (lQCD) \cite{Li:2017eic,Gao:2021xsm}. For comparison, the experiments report \cite{ParticleDataGroup:2022pth}: $\pi: 0.659(4)$, $K_\pm: 0.560(31)$,  $K_0: 0.277(18)i$. The units are fm. }
\end{table}

For the full charge radius of these mesons, each approaches reports results of the same order of magnitude, that is, $   \sqrt{\left\langle r^2\right\rangle} < 1$ fm. It is no surprise because fm is the order of magnitude of a nucleon's size. Nevertheless, the charge radii obtained from contact interaction model (CI) are lower than others, since CI is a relatively  simple model \cite{Hernandez-Pinto:2023yin}. However, it still provides useful qualitative results. CI predicts that the charge radius of $\pi$ is the largest of these ground state pseudo-scalar meson, and our results confirm this conclusion. This is interesting because $\pi$ is the one with the most significant DCSB effect, which seems to suggest that DCSB, while generating mass, also tends to increase the size of the system. \par 
For slightly/moderately flavor asymmetric meson, such as $u\bar{s}$, $u\bar{c}$, $s\bar{c}$, $s\bar{b}$, $c\bar{b}$, ..., the charge radii predicted by these methods are generally consistent. However, in the case of extremely flavor symmetry breaking, such as $u\bar{b}$, these results begin to deviate from each other. CI reports 0.34 fm for $B^+$, but the result of AM is 0.926 fm, even larger than the proton radius, 0.841(1) fm \cite{ParticleDataGroup:2022pth}. In this work, the predicted charge radius of $B^+$ is rather close to light-front framework (LFF) $\sim 0.62$ fm. More comparison can be found in Table. \ref{tab:PS}. \par
\begin{figure}
\centering 
\includegraphics[width=.49\textwidth]{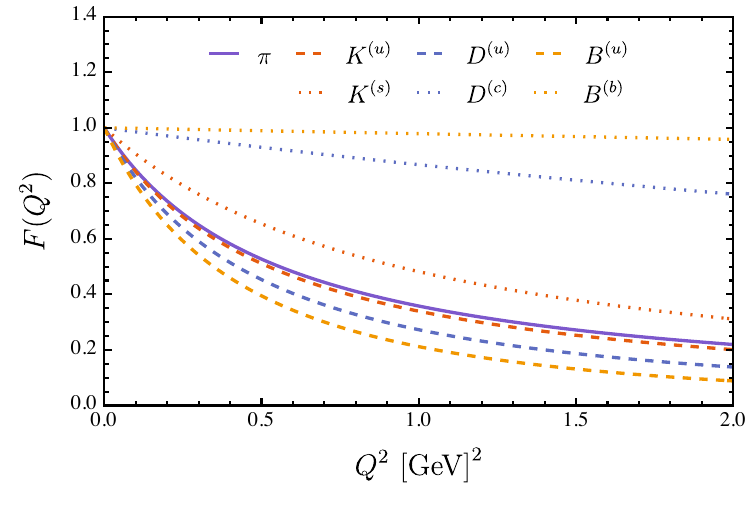}
\includegraphics[width=.49\textwidth]{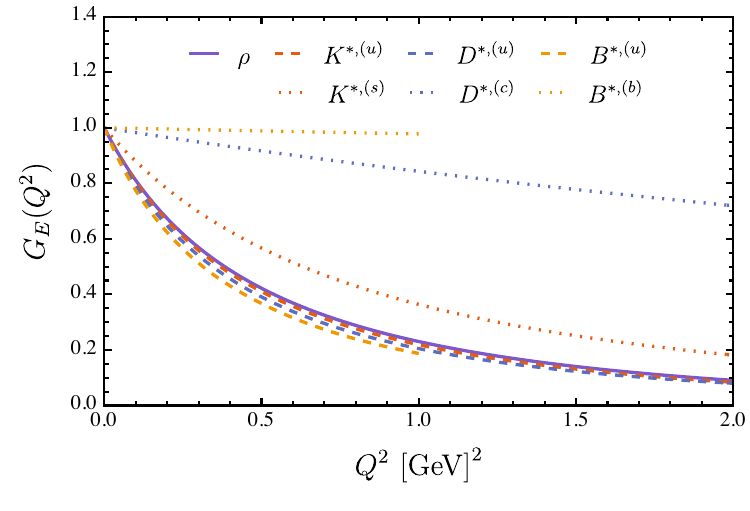}
\includegraphics[width=.49\textwidth]{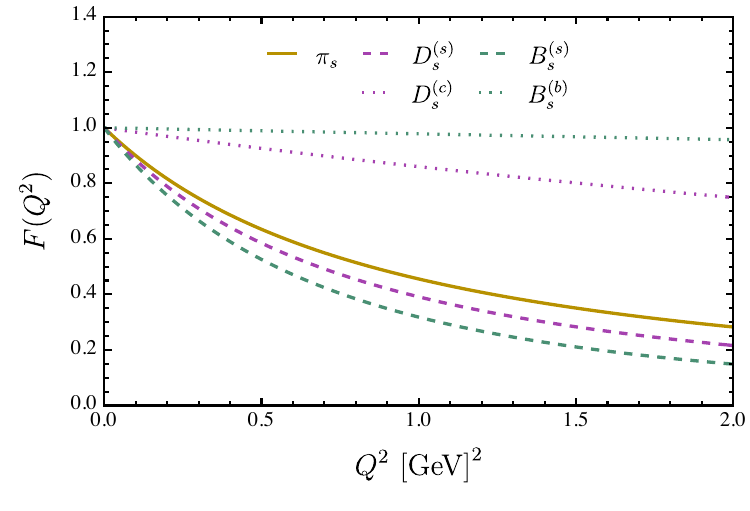}
\includegraphics[width=.49\textwidth]{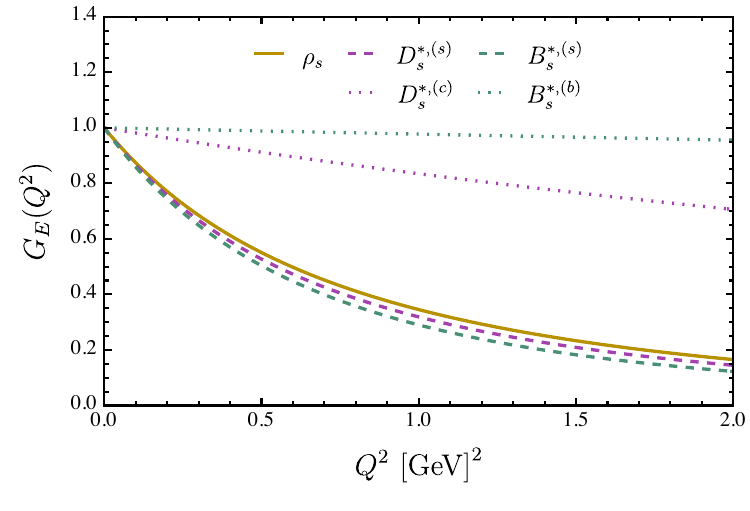}
\includegraphics[width=.49\textwidth]{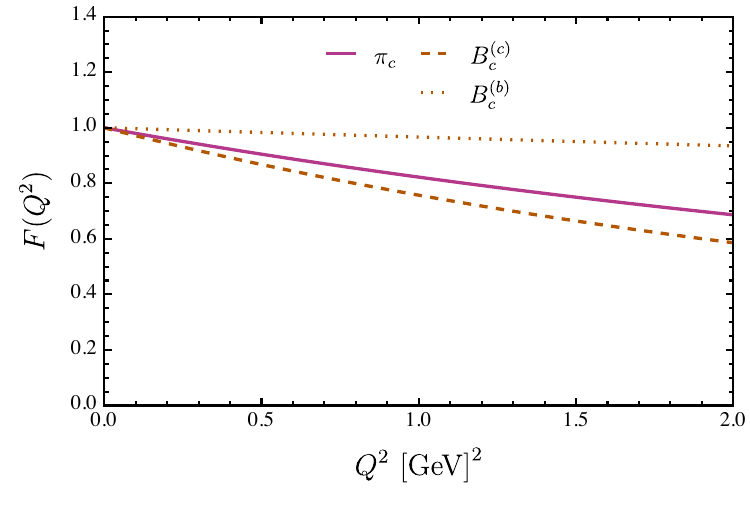}
\includegraphics[width=.49\textwidth]{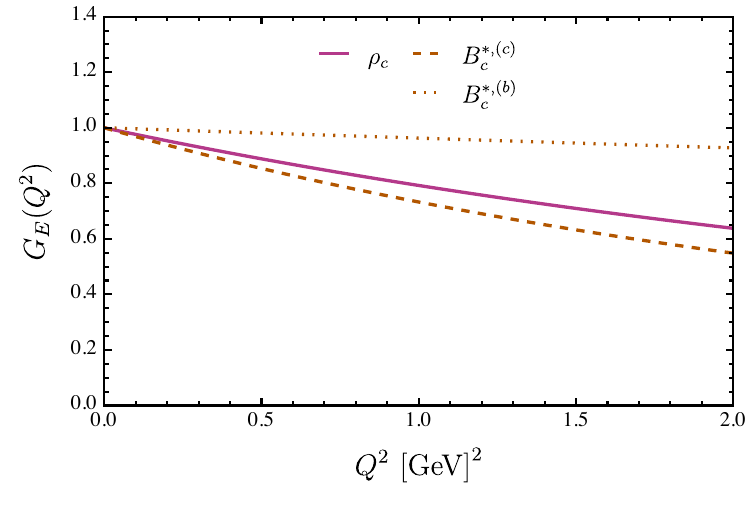}
\includegraphics[width=.49\textwidth]{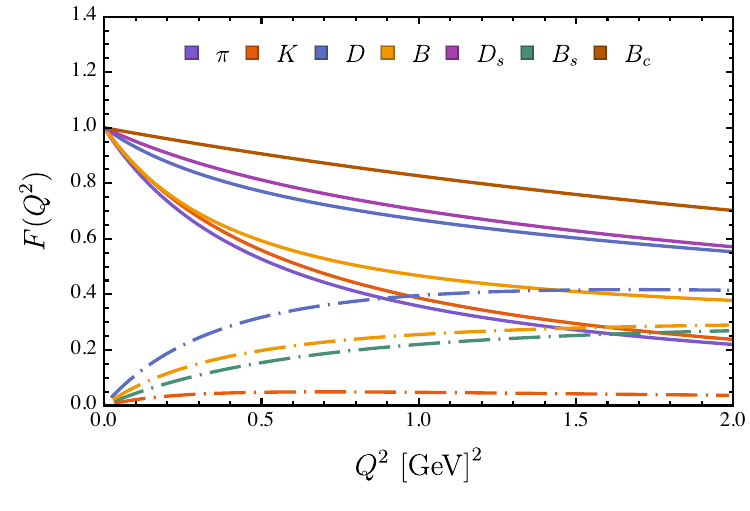}
\includegraphics[width=.49\textwidth]{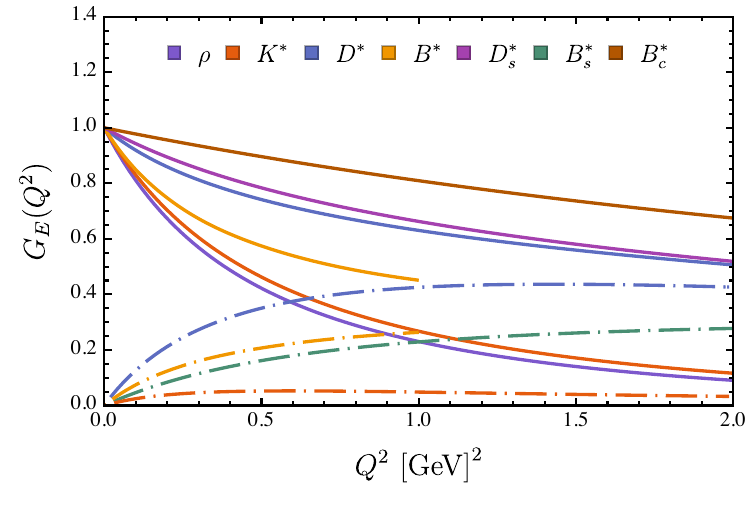}
\caption{\label{fig:G_E} The electric form factor of heavy-light pseudo-scalar/vector mesons. Left panel: pseudo-scalar mesons, right panel: vector mesons. Solid lines: the results of charged mesons; dot-dashed lines: the results of neutral mesons; dashed/dotted lines: the contribution of lighter/heavier (anti-)quark in the system (see Eq (\ref{eq.eff2})), where $K^{(u)}$ means $u$ (anti-)quark's contribution in $K$ meson, and other legends are similar.}
\end{figure}

 
According to Eq. (\ref{eq:eff}), the contributions of different quarks to the system can be extracted, this gives us a glimpse into the internal structure of heavy-light meson. In Table. \ref{tab:PS} and Figure. \ref{fig:G_E},  both the separated contributions and full results are presented, we will discuss this in conjunction with vector mesons in the next subsection.

\subsection{Vector mesons}

Compared with the pseudo-scalar case, the heavy-light vector meson's electromagnetic form factor have received much less attention. On the one hand, it is difficult to measure in experiment, on the other hand, its theoretical calculation is more complicated. However, noteworthy distinctions or unexpected similarities between pseudo-scalar and vector mesons, such as their charge radii, can help us to better understand the internal structure of these hadrons. Therefore, in this work, we calculate the EFFs of pseudo-scalar/vector mesons with momentum transfer $Q^2 < 2 \ \text{GeV}^2$ uniformly, except $B^*$ because the pole of the quark propagator in the complex plane limits the computable region \cite{Chen:2018rwz}.\par
\begin{table}
\centering
\scalebox{0.8}{
\begin{tabular}{|l|l|lll|lll|lll|}
\hline
\multicolumn{2}{|c|}{Meson}&\multicolumn{3}{|c|}{$\sqrt{\langle r^2 \rangle}$} & \multicolumn{3}{|c|}{$\mu$} & \multicolumn{3}{|c|}{$\mathcal{Q}$}\\
\multicolumn{2}{|c|}{}   & \multicolumn{1}{c}{L} & \multicolumn{1}{c}{H} & \multicolumn{1}{c|}{Full}   & \multicolumn{1}{c}{L} & \multicolumn{1}{c}{H} & \multicolumn{1}{c|}{Full}    & \multicolumn{1}{c}{L} & \multicolumn{1}{c}{H} & \multicolumn{1}{c|}{Full}  \\ \hline
 $\rho$&$u\bar{d}$ & 0.722 & 0.722 & 0.722 & 2.006 & 2.006 & \ 2.006 & -0.364 & -0.364 & -0.364\\ 
 $K^{*+}$&$u\bar{s}$ & 0.732 & 0.557 & 0.679 & 2.234 & 1.896 & \ 2.121 & -0.479 & -0.340 & -0.433\\
 $K^{*0}$&$d\bar{s}$ & / & / & 0.274$i$ & / & / & -0.112 & / & / & \ 0.046\\
 $D^{*+}$&$c\bar{d}$ & 0.768 & 0.204 & 0.473 & 4.464 & 1.419 & \ 2.434 & -1.281 & -0.148 & -0.525\\
 $D^{*0}$&$c\bar{u}$ & / & / & 0.604$i$ & / &/ & -2.030 & / & / & \ 0.755\\
 $B^{*+}$&$u\bar{b}$ & 0.802& 0.072 & 0.657 & 11.23 & 1.178 & \ 7.880 & -3.238 & -0.056 & -2.177\\
 $B^{*0}$&$d\bar{b}$ & / & / & 0.462$i$ & / & / & -3.351 & / & / & \ 1.061\\
 $D^*_s$&$c\bar{s}$ & 0.589 & 0.210 & 0.381 & 3.856 & 1.472 & \ 2.267 & -1.044 & -0.170 & -0.461\\
 $B^*_s$&$s\bar{b}$ & 0.605 & 0.074 & 0.347$i$ & 9.526 & 1.204 & -2.774 & -2.719 & -0.059 & \ 0.887\\
 $B^*_c$&$c\bar{b}$ & 0.275 & 0.094 & 0.231 & 4.078 & 1.424 & \ 3.193 & -0.945 & -0.118 & -0.669\\
  \hline
\end{tabular}
}
\caption{\label{tab:VC} The charge radius, $\sqrt{\langle r^2 \rangle}$, magnetic moment, $\mu$, and the quadrupole moment, $\mathcal{Q}$, of heavy-light vector meson, including the contributions of lighter quark (L) and heavier quark (H). The units are fm, $e / 2 M_V$, $e / M_V^2$, respectively. For  comparison, an experimental values of $\sqrt{\langle r^2 \rangle_\rho}$ is 0.721(35) \cite{Povh:1990ad}, and Ref. \cite{Bhagwat:2006pu} reports $\sqrt{\langle r^2 \rangle_\rho}=0.73$ fm, $\mu_\rho=2.01$, $Q_\rho= -0.41$, $\sqrt{\langle r^2 \rangle_{K^*}}=0.656$ fm, $\mu_{K^*}=2.23$, $Q_{K^*}= -0.38$, $\sqrt{\langle r^2 \rangle_{K^{*0}}}=0.282i$ fm, $\mu_{K^{*0}}=-0.26$, $Q_{K^{*0}}= 0.01$ under RL kernel, with $\sqrt{\langle r^2 \rangle_\rho}=0.748(27)$ fm from another model \cite{Krutov:2016uhy}.}
\end{table}

\begin{table}
\centering
\scalebox{0.8}{
\begin{tabular}{|l|l|llllllll|}
\hline
\multicolumn{2}{|c|}{Meson} & \multicolumn{1}{c}{This work} &  \multicolumn{1}{c}{ENJL} & \multicolumn{1}{c}{LCSR} & EBM & BM & NR & BSLT & ChPT \\ \hline
 $\rho$&$u\bar{d}$ &\ 2.492 & 2.54 & - & \ 2.500 & - & - & - & -\\ 
 $K^{*+}$&$u\bar{s}$ &\ 2.261 & 2.26 & - & \ 2.210 & - & - & - & - \\
  $K^{*0}$&$d\bar{s}$ & -0.119 & - & - & -0.216 & - & - & - & -  \\
 $D^{*+}$&$c\bar{d}$ & \ 1.132 & 1.16 & \ 1.16(8) &\ 1.060 & \ 1.17 & \ 1.32 & - & \ $1.62_{-0.08}^{+0.24}$\\
  $D^{*0}$&$c\bar{u}$ & -0.944 & - & \ 0.30(4) & -1.210 &-0.89 & -1.47 & - & -$1.48_{-0.38}^{+0.22}$ \\
 $B^{*+}$&$u\bar{b}$ &\ 1.386 & 1.47 & \ 0.90(19) & \ 1.470 &\ 1.54 & \ 1.92 & - &\ $1.77_{-0.30}^{+0.25}$\\
  $B^{*0}$&$d\bar{b}$ & -0.589 & - & -0.21(4) & -0.650 & -0.64 & -0.87 & - & -$0.92_{-0.11}^{+0.15}$ \\
 $D^*_s$&$c\bar{s}$ & \ 1.007 & 0.98 & \ 1.00(14)& \ 0.870 & \ 1.03 & \ 1.00 & - & \ $0.69_{-0.10}^{+0.22}$\\
 $B^*_s$&$s\bar{b}$ & -0.480 & - & -0.17(2)& -0.480 & -0.47 & -0.55 & - & -$0.27_{-0.10}^{+0.13}$\\
 $B^*_c$&$c\bar{b}$ & \ 0.472 & - &- & \ 0.350 &\ 0.56 & \ 0.45 & 0.426 & -\\
  \hline
\end{tabular}
}
\caption{\label{tab:VC.mu2} Magnetic moments (in nuclear magneton) of heavy-light vector mesons. The results of other approaches come from: Extended Nambu–Jona-Lasinio model (ENJL) \cite{Luan:2015goa}, light cone sum rules (LCSR) \cite{Aliev:2019lsd}, Extended-Bag model (EBM)\cite{Simonis:2016pnh}, Bag model(BM) \cite{Bose:1980vy}, non-relativistic quark model (NR) \cite{Simonis:2016pnh,ParticleDataGroup:2018ovx}, Blankenbecler-Sugar equation (BSLT) \cite{Lahde:2002wj} and chiral
perturbation theory (ChPT) \cite{Wang:2019mhm}.}
\end{table}

 The electric form factor of vector meson, $G_E(Q^2)$, can be compared with the pseudo-scalar meson's $F(Q^2)$, because both of them correspond to the charge distribution. The results of this comparison are presented in Figure. \ref{fig:G_E}, and a clear pattern can be noticed immediately. \par 
As mentioned earlier, Eq. (\ref{eq:eff}) allows us to separate the contributions of the lighter and heavier dressed-quark. For $u\bar{q}$, $q=u/d,s,c,b$ systems, with the increase of current quark mass of $q$, the form factor of the lighter quark becomes steeper while the form factor of the heavier quark flattens out (see Figure. \ref{fig:G_E}, upper panel). The extracted charge radii are presented in the Table. \ref{tab:PS} (PS) and Table. \ref{tab:VC} (VC). Obviously, the distribution range of $u$ and $\bar{q}$ (anti-)quarks gradually expands and contracts. Especially, for $u$ quark in  $u\bar{b}$ system, $\sqrt{2/3\langle r^2 \rangle_u} = $ 0.618 fm (PS), 0.655 fm (VC), it is very close to the charge radius of $B^+$ and $B^{*+}$, that is, 0.619 fm and 0.657 fm. This reveals that, in $u\bar{b}$ systems, the $\bar{b}$ quark is basically stationary, with the charge distribution almost entirely contributed by the $u$ quark. \par 
For the $s\bar{q}$, $c\bar{q}$ systems, in the middle panel of Figure. \ref{fig:G_E} we constructed four fictitious charged states, $\pi_s,\pi_c,\rho_s,\rho_c$, which are constituted from $q=u/d$-like quarks with current masses equal to  $s$ and $c$ quarks. Because in the definition of this work, the electromagnetic form factor of  neutral  $s\bar{s}$, $c\bar{c}$ system is trivially zero (see, Eq. (\ref{eq.eff2})). Again, flavor symmetry breaking leads to the splitting of the form factor, the distribution range of lighter and heavier quarks gradually expands and contracts, respectively. We note that similar conclusions have also been reported by the ENJL model \cite{Luan:2015goa}. \par 
The full results of the electric form factors of heavy-light mesons are presented in the lower panel of Figure. \ref{fig:G_E}. Qualitatively, the results of pseudo-scalar and vector channel are not much different, however, the charge radius of vector meson is larger than that of its pseudo-scalar counterpart (see Table. \ref{tab:PS} and Table. \ref{tab:VC}). This suggests that spin-dependent interactions will expand the size of the meson, which conforms to the conclusion in the case of flavor symmetry \cite{Xu:2019ilh}.\par 

\begin{figure}
\centering 
\includegraphics[width=.49\textwidth]{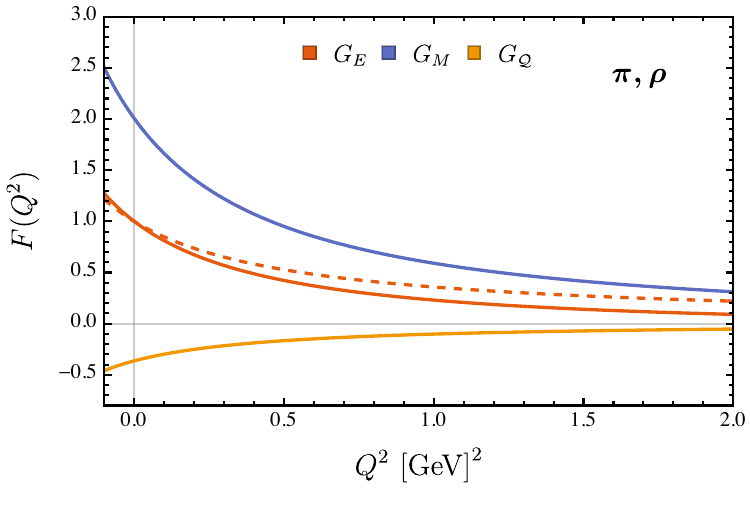}
\includegraphics[width=.49\textwidth]{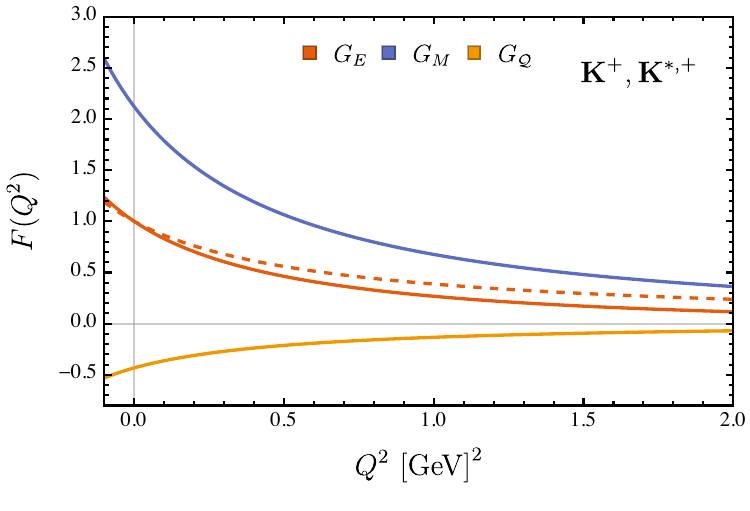}
\includegraphics[width=.49\textwidth]{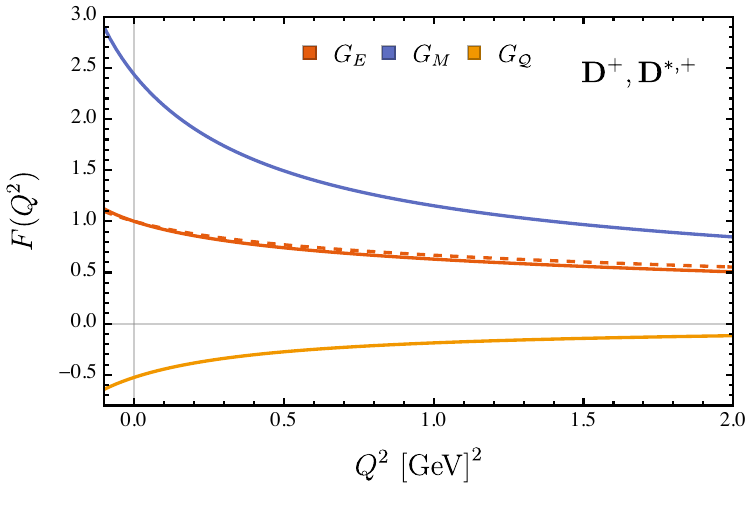}
\includegraphics[width=.49\textwidth]{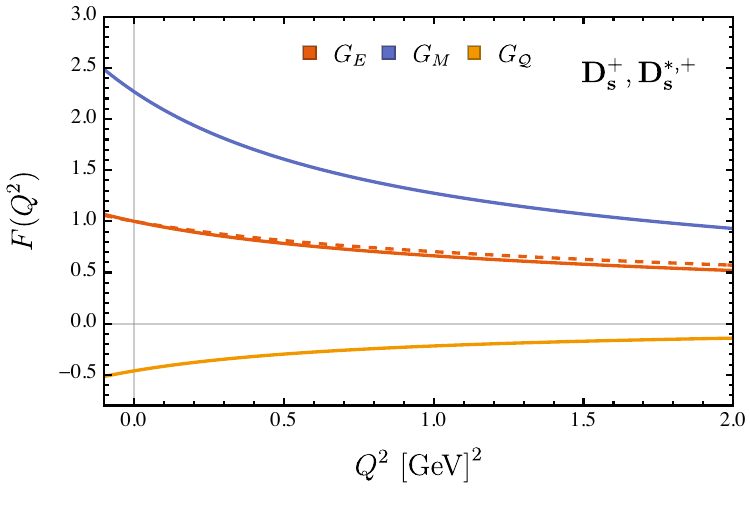}
\includegraphics[width=.49\textwidth]{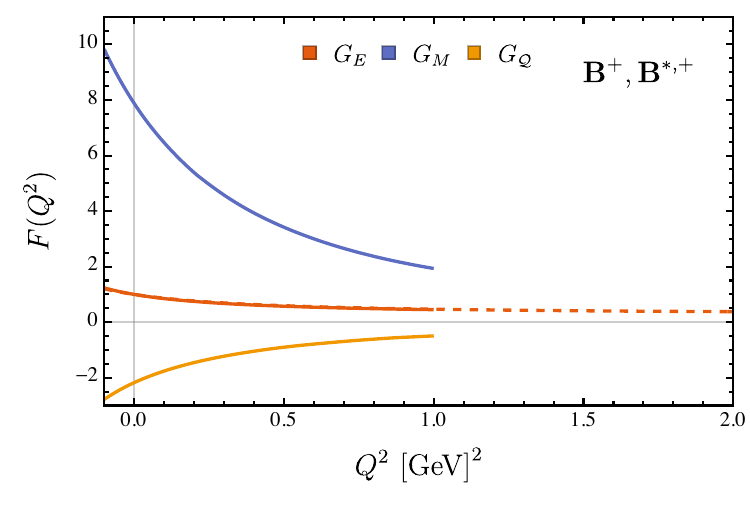}
\includegraphics[width=.49\textwidth]{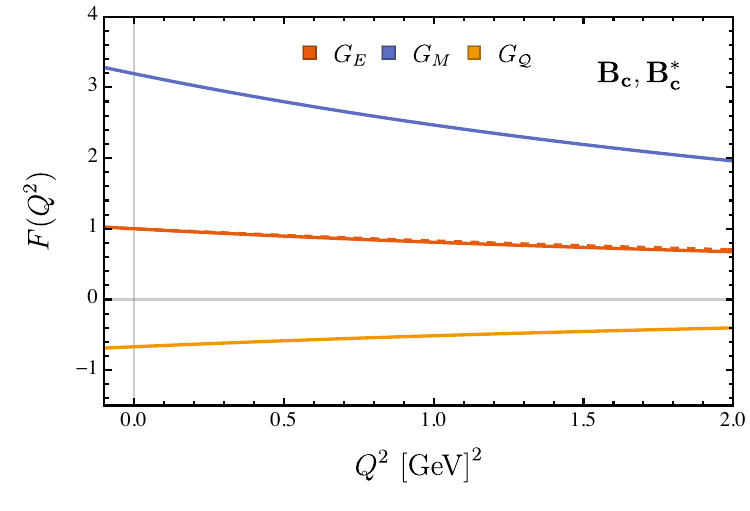}
\caption{\label{fig:c} The electromagnetic form factors of heavy-light charged mesons. Solid lines: vector mesons; dashed lines: pseudo-scalar mesons.}
\end{figure}

\begin{figure}
\centering 
\includegraphics[width=.49\textwidth]{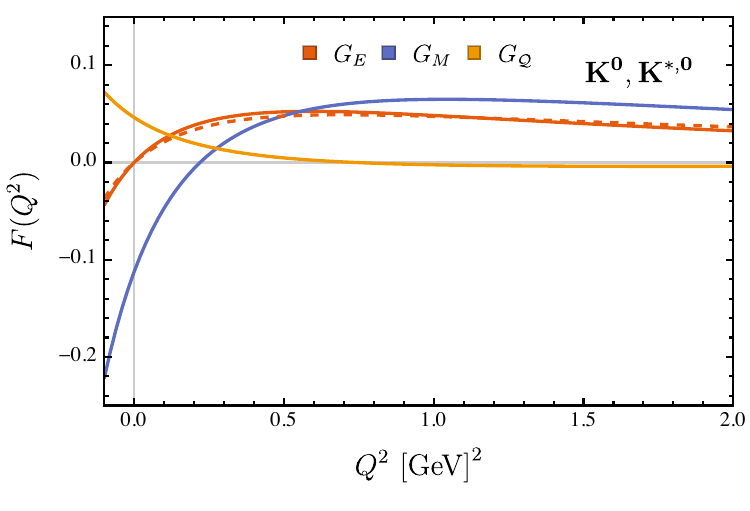}
\includegraphics[width=.49\textwidth]{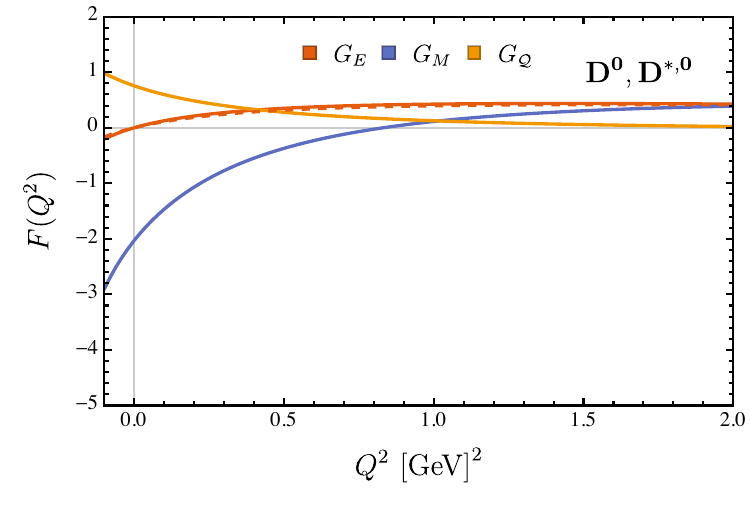}
\includegraphics[width=.49\textwidth]{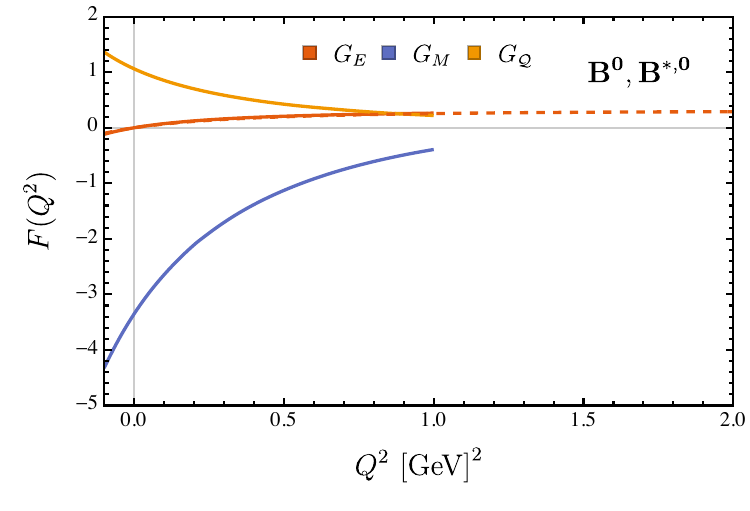}
\includegraphics[width=.49\textwidth]{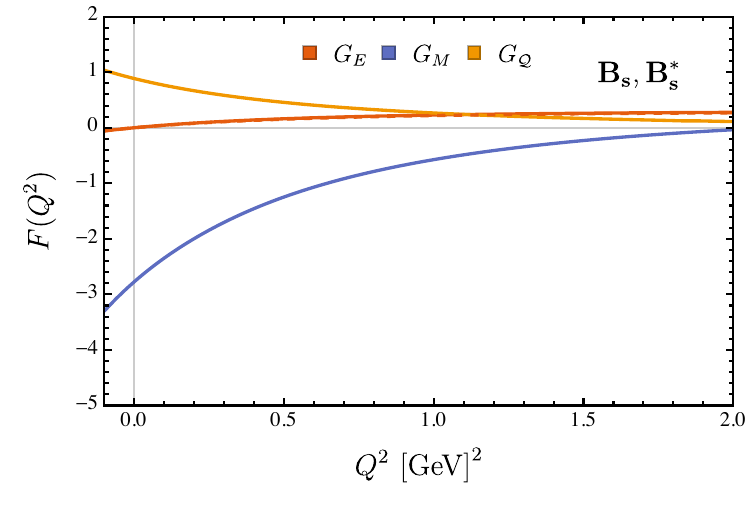}
\caption{\label{fig:n} The electromagnetic form factors of heavy-light neutral mesons. Solid lines: vector mesons; dashed lines: pseudo-scalar mesons.}
\end{figure}

Differing from pseudo-scalar mesons, vector mesons have two more form factors, $G_M(Q^2)$ and $G_{\mathcal{Q}}(Q^2)$,  corresponding to magnetic moment and quadrupole moment, respectively. The full results are plotted in Figure. \ref{fig:c} (charged mesons)  and Figure. \ref{fig:n} (neutral mesons). For heavy-light mesons, the form factor is qualitatively consistent as the flavor asymmetry increases. However, in the case of charge neutrality, the significant deviation from zero has been exhibited, which reveals a non-trivial internal structure. \par 
 The extracted magnetic moments and quadrupole moments are presented together in Table. \ref{tab:VC}. It is easy to see that when the charge radius of dressed-quark rises, the corresponding magnetic moment and quadrupole moment also increases. In Table. \ref{tab:VC.mu2}, the magnetic moments are listed in the unit of nuclear magneton $\mu_n$ to compare with the results of other approaches. Once again, the results given by each model/framework are basically the same for slightly/moderately flavor asymmetric meson, but in the extremely flavor asymmetric case, there are significant differences between them. Our predictions of the magnetic moment are basically consistent with the ENJL model's results. The possible reason is that NJL is a similar framework to DSEs/BSEs, and the magnetic moment is not sensitive to the form of interaction. However, due to the absence of experimental data, these predictions still need to be verified by more approaches.
\section{Summary}
\label{sec:4}

In this work,  we systematically investigate the electromagnetic form factors of heavy-light pseudo-scalar/vector mesons within DSEs/BSEs framework for the first time, including $u\bar{s}$, $u\bar{c}$, $u\bar{b}$, $s\bar{c}$, $s\bar{b}$ and $c\bar{b}$ systems. Based on it, we extract the charge radii of pseudo-scalar mesons, the charge radii, magnetic moments and the quadrupole moments of the vector mesons, then compare our results with those obtained by other approaches. \par 
The numerical results show that, the flavor symmetry breaking will lead to a splitting of the form factor of different quark in heavy-light system, and the distribution range of lighter and heavier quark gradually expands and contracts, respectively. In the case of vector meson, when the charge radius of dressed-quark increases, so do its corresponding magnetic moment and quadrupole moment. The competition between the contributions of the lighter and heavier quark together generates the electromagnetic form factors of heavy-light meson. \par 
The results presented in this work can be compared with the experimental data and further theoretical calculations, such as applying more elaborate beyond-RL kernel, in the future. We expect that it will be useful for the understanding of the internal structure and dynamics of QCD's bound states.

\acknowledgments
We would like to thank Jorge Segovia, Khépani Raya, José Rodríguez-Quintero, Craig D. Roberts for useful discussions/suggestions. This work has been partially funded by Ministerio Espa\~nol de Ciencia e Innovaci\'on under grant Nos. PID2019-107844GB-C22 and PID2022-140440NB-C22; Junta de Andaluc\'ia under contract Nos. Operativo FEDER Andaluc\'ia 2014-2020 UHU-1264517, P18-FR-5057 and also PAIDI FQM-370. The authors acknowledge, too, the use of the computer facilities of C3UPO at the Universidad Pablo de Olavide, de Sevilla.

\bibliography{ref.bib}
\bibliographystyle{JHEP.bst}

\end{document}